\documentclass{PoS}
\usepackage{amsmath,amssymb,epsfig}

\newcommand{\mathe}{\mathrm{e}}
\newcommand{\tmop}[1]{\ensuremath{\operatorname{#1}}}
\newcommand{\be}{\begin{equation}}
\newcommand{\ee}{\end{equation}}

\title{Cluster simulation of two-dimensional relativistic fermions}

\ShortTitle{Cluster simulation of two-dimensional relativistic fermions}

\author{\speaker{Ulli Wolff}\\
        Institut f\"ur Physik, Humboldt Universit\"at\\ 
        Newtonstr. 15 \\ 
        12489 Berlin, Germany\\
        E-mail: \email{uwolff@physik.hu-berlin.de}}

\abstract{The (discrete) Gross-Neveu model is studied in a lattice realization with
an $N$-component Majorana Wilson fermion field. It has an internal O($N$) symmetry in addition
to the euclidean lattice symmetries. The
discrete chiral symmetry for vanishing mass is expected to emerge
in the continuum limit only.
The lattice theory is first recast in terms of two-valued bosonic link variables (dimers).
In this representation, which coincides with the loop representation obtained earlier
by Gattringer with the help of eight-vertex-models, the Boltzmann weight is essentially
positive. 
While standard local updates are
possible in this form we construct a further 
exact transformation where we generate dimer configurations
as Peierls contours of an Ising model with a local action residing
on plaquettes. For this model a Swendsen-Wang type cluster algorithm is constructed.
At vanishing coupling it is numerically demonstrated to almost completely eliminate
critical slowing down. Although further tests are required, an avenue to numerical studies
of the Gross-Neveu model with unprecedented precision seems open.
\vspace{3cm}
\begin{flushright}
HU-EP-07/54\\
SFB/CPP-07-65
\end{flushright}
}

\FullConference{The XXV International Symposium on Lattice Field Theory\\
		 July 30-4 August 2007\\
		 Regensburg, Germany}

\begin{document}

\section{Introduction}
The O($N$) symmetric Gross-Neveu (GN) model \cite{Gross:1974jv} forms a particularly
simple  family of self-interacting
fermionic quantum field theories. It may be regarded as the natural fermionic counter-part
of the [$N$-component] $\phi^4$ theory of real bosons. While they are renormalizable in
four dimensions the same is true in $d=2$ for GN, where it is even asymptotically free.
It is most naturally written in terms of Majorana fermions $\xi$ which are pairwise
related to the more familiar Dirac fermions by
\be
   \psi = \frac{1}{\sqrt{2}} (\xi_1 + i \xi_2), \hspace{1em} \overline{\psi} =
   \frac{1}{\sqrt{2}} (\xi_1^{\top} - i \xi_2^{\top})\mathcal{C}, \hspace{1em}
   \text{$\mathcal{C}$} = - \text{$\mathcal{C}$}^{\top} : \tmop{charge}
   \tmop{conjugation}.
\ee
We here define the action immediately on the lattice with Wilson-fermions as
\be
   S = \sum_x \left\{  \frac{2 + m}{2} \xi^{\top}_{} \mathcal{C} \xi -
   \frac{g^2}{8} ( \xi^{\top} \mathcal{C} \xi)^2 \right\} - \sum_{x \mu}
   \left\{ \xi^{\top} (x)\mathcal{C}\, \frac{1-\gamma_\mu}{2}\,  \xi (x + \hat{\mu})
   \right\}.
\label{SGN}
\ee
We work with lattice units, $a=1$. The first term contains the mass and the diagonal
part of the Wilson term and is followed by the interaction and the hopping term.
Note that for Majorana fermions the backward hopping terms coincide the forward ones.
All $\xi$-bilinears are contracted in the internal $N$-valued index making the 
O($N$) symmetry manifest. It allows only one type of 4-fermi interaction, hence
the strict renormalizability.
The discrete chiral transformation $\xi \to \gamma_5 \xi$ is broken only by the mass
and the Wilson term. As with chiral symmetry in QCD it is expected to emerge in the massless
continuum limit where the bare Wilson mass has to be tuned to a critical value $m_c$.

The partition function follows,
\be
   Z_{\tmop{GN}} = \int D \xi \mathe^{- S} = \exp \left\{ \frac{g^2}{2} \sum_x
   \frac{\partial^2}{\partial m (x)^2} \right\} (Z_0 [m (x)])^N.
\ee
By allowing for an $x$-dependent mass $m$ we re-wrote it here in terms of the partition
function $Z_0$ of {\em one} Majorana fermion in such a background. Its action is the bilinear part
of (\ref{SGN}). In ref.\cite{Wolff:2007ip} more details on everything said here can be found.
In particular it is explained, how the interacting theory can be simulated once one has the
efficient algorithm for $Z_0[m(x)]$ available that we study below. Such simulations are
presently under way.

\section{Transformation to a dimer ensemble}
\begin{figure}
\begin{center}
\includegraphics[width=0.8\textwidth]{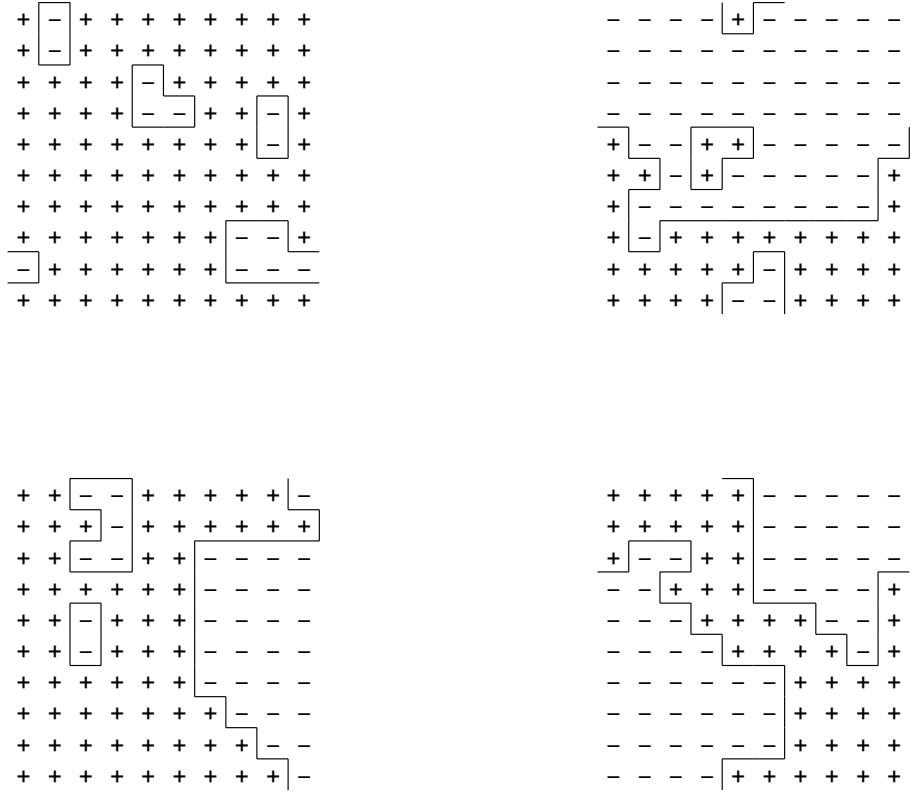}
\caption{Lines correspond to dimer loops which also form contours
separating domains of up and down spins. The four different possible loop topologies
on the torus are exemplified.}
\label{classes}
\end{center}
\end{figure}

To expand the partition function with $\varphi(x) = 2 + m(x)$ we first exploit the nilpotency
of even Grassmann elements and obtain.
\be
   Z_0 = \int D \xi \prod_x \left\{ 1 + \varphi \xi_2 \xi_1 \right\} 
   \prod_{x, \mu} \left( 1 + \xi^{\top}_{} (x)\mathcal{C}\frac{1-\gamma_\mu}{2}\, \xi (x
   + \hat{\mu}) \right).
\ee
Next the product of the hopping terms on each link is organized with the help
of link dimer-variables $k(x,\mu)=0,1$,
\begin{eqnarray}
  Z_0 & = & \sum_{\{k (x, \mu)\}} \int D \xi \prod_x \left\{ 1 + \varphi \xi_2
  \xi_1 \right\} \prod_{x, \mu} \left( \xi^{\top}_{} (x)\mathcal{C}\frac{1-\gamma_\mu}{2}\,
  \xi (x + \hat{\mu}) \right)^{k (x, \mu)}\\
  & = & \sum_{\{k (x, \mu)\}} \rho [k] \times \tmop{sign} [k]
\label{Zkdef}
\end{eqnarray}
with
\be
    \rho [k] = \prod_x \times \left\{ \begin{array}{ll}
     \varphi (x) & \tmop{if} \tmop{no} \tmop{dimer} \tmop{at} x (\equiv
     \tmop{monomer})\\
     1 & \tmop{if} 2 \tmop{dimers} \tmop{in} \tmop{the} \tmop{same}
     \tmop{direction} \tmop{at} x\\
     1 / \sqrt{2} & \tmop{if} 2 \tmop{dimers} \tmop{in} \tmop{different}
     \tmop{directions} \tmop{at} x\\
     0 & \tmop{else}
   \end{array} \right. .
\label{rhovals}
\ee
To arrive at the weight $\rho$ one has to observe that 
the Grassmann integral constrains dimers (links with $k(x,\mu)=1$) to
form closed loops and at each site either two or zero dimers can touch. A site with no dimer adjacent
is called a monomer and leads to a factor $\varphi(x)$. The weight for such loops can be
computed analytically \cite{Stamatescu:1980br} and leads to (\ref{rhovals}). The factor sign$[k]$ equals $+1$
unless dimer loops wind around the boundary, in which case it depends on whether the boundary
conditions are periodic or antiperiodic. This leads to relations between fermion and dimer-ensembles
of the type
\be
   Z_{\xi}^{- +} = \text{$Z_k^{00}$+$Z_k^{10} -
   Z_k^{01}$+$Z_k^{11}$} \hspace{1em} (+ 3 \tmop{more} \tmop{such}
   \tmop{relations}).
\label{Zxifix}
\ee
In this formula $Z_{\xi}^{- +}$ is a fermion (anti)periodic in (time) space, and $Z_k^{10}$
is a partition function of the type (\ref{Zkdef}) restricted to configurations with 1 (mod 2) loops
around the torus in time and 0 in space and analogously for the other cases. The four possibilities
are illustrated by the lines in Fig.\ref{classes}.
An example of the translation of observables between the representations,
derived by taking derivatives with respect to $m(x)$, relates the scalar fermion
density and the monomer density
\be
   - \frac{2 + m}{2}  \left\langle \xi^{\top}_{} \mathcal{C} \xi
   \right\rangle_{\tmop{fermion}} = \left\langle K (x)
   \right\rangle_{\tmop{dimer}}, \hspace{1em} K (x) = \left\{
   \begin{array}{lll}
     1 & \tmop{if} & \tmop{monomer} \tmop{at} x\\
     0 &  & \tmop{else}
   \end{array} \right. .
\ee

\section{Spin representation and cluster algorithm}
In the form of the dimer ensembles we have recast the originally fermionic theory
in terms of discrete commuting variables with local interactions. They can be updated
by a local Metropolis algorithm as demonstrated in \cite{Gattringer:2007em}. Due to the
constraints [the zero in (\ref{rhovals})] the minimal updates have to change dimers
around a plaquette. Such updates do not change the dimer loop topology which is
determined by the initial state. Such a fixed topology dimer ensemble corresponds to a combination of
fermion ensembles with several kinds of boundary conditions. The local simulation of
such an ensemble has roughly the same complexity as simulating a standard Ising model.

\begin{figure}
\begin{center}
\includegraphics[width=0.8\textwidth]{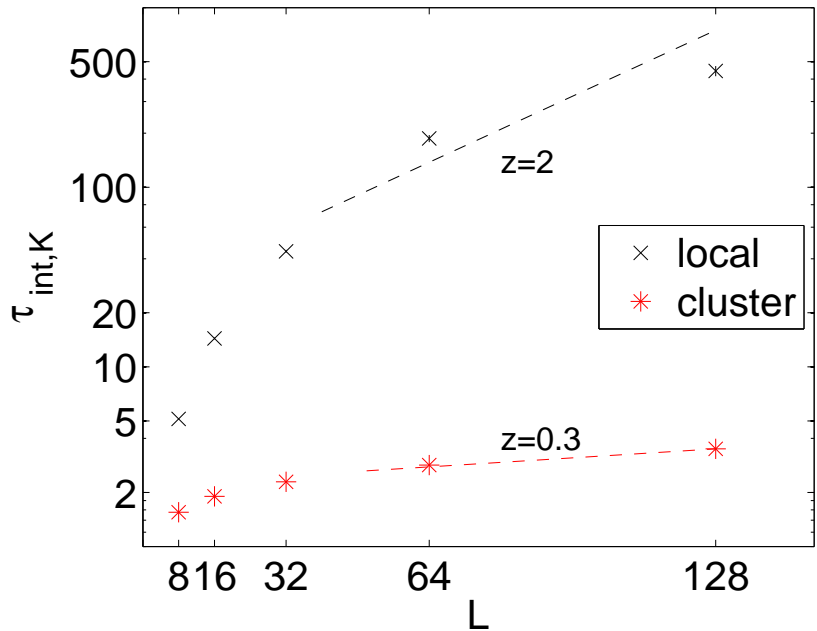}
\caption{Autocorrelation times for local ($\times$) and cluster ($\ast$)
algorithm for a massless fermion.}
\label{taufig}
\end{center}
\end{figure}

Also as in the Ising model the local update leads to critical slowing down with a
dynamical exponent $z$ around two. For the case equivalent to a free Wilson Majorana fermion at
vanishing mass, which is critical, the upper crosses in Fig.\ref{taufig} 
at least qualitatively show the expected
growth of the integrated autocorrelation time for the monomer density.
In the Ising model this severe deterioration of efficiency is practically
completely eliminated by collective cluster update procedures like the one
of Swendsen and Wang \cite{Swendsen:1987ce}.
The good news is now that also here, for our Ising type theory
with plaquette interaction -- exactly equivalent to Wilson fermions --
we have found an equally efficient technique. We briefly sketch it here,
but for a more detailed understanding probably \cite{Wolff:2007ip} has to be consulted.

We consider the lattice dual to the one carrying the dimers (and originally the fermions).
Its sites can be drawn in the centers of the original plaquettes. This is where we put
Ising spins, which appear as plus and minus signs in Fig.\ref{classes}. 
We also read off there, how the closed (non-intersecting) dimer loops are obtained as
the Peierls contours encircling areas of a given spin orientation. As discussed in
\cite{Wolff:2007ip} this is an exact two-to-one mapping of Ising configurations
obeying a certain plaquette constraint and allowed dimer configurations. The four
different dimer topologies correspond to the four possible combinations
of (anti)periodic boundary
conditions of the Ising spins. There antiperiodicity forces domain boundaries
into our `magnet' which are precisely the non-contractable dimer loops,
see again Fig.\ref{classes}. 
The Ising partition function
-- for one choice of topology/boundary conditions ---
is now given by
\be
   2 Z_k^{10} =  Z_s^{+ -} =  \sum_{\{s (
   \underline{x})\}} \prod_{\tmop{plaquettes}} w
   \left(\begin{array}{cc}
     s_4 & s_3\\
     s_1 & s_2
   \end{array}\right).
\ee
The desired weights for the individual configurations dictate the values
of $w$,
\be
   w \left( \begin{array}{l}
     \resizebox{0.7cm}{!}{\epsfig{file=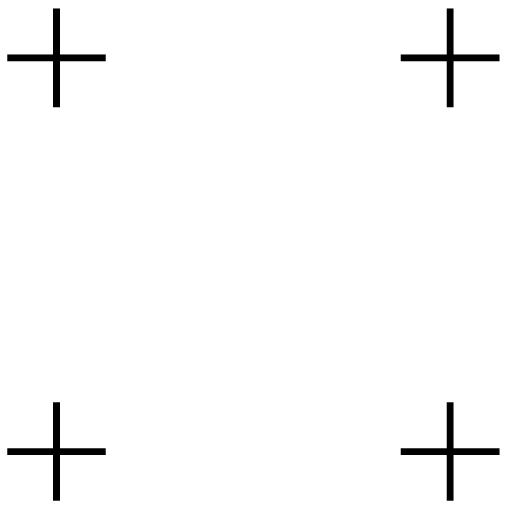}}
   \end{array} \right) = \varphi (x), \hspace{1em} w \left( \begin{array}{l}
     \resizebox{0.7cm}{!}{\epsfig{file=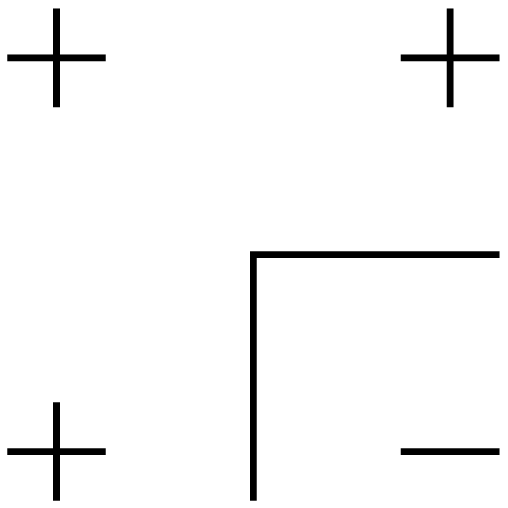}}
   \end{array} \right) = \frac{1}{\sqrt{2}}, \hspace{1em} w \left(
   \begin{array}{l}
     \resizebox{0.7cm}{!}{\epsfig{file=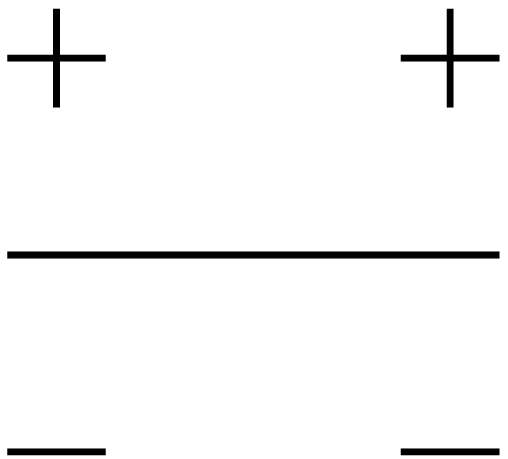}}
   \end{array} \right) = 1, \hspace{1em} w \left( \begin{array}{l}
     \resizebox{0.7cm}{!}{\epsfig{file=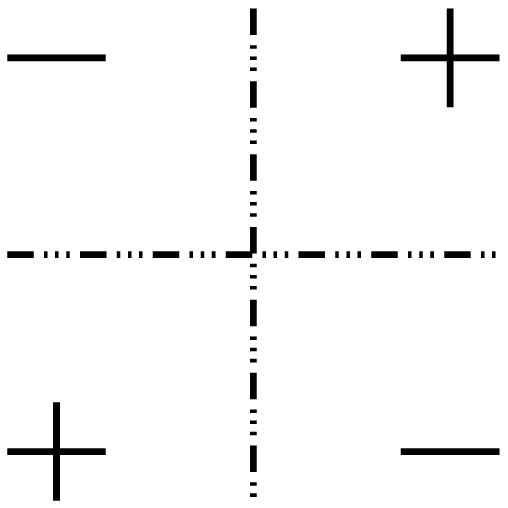}}
   \end{array} \right) =  0, \ldots . 
\ee
with other values obtained form these by rotating or flipping the four spins.
The essential observation for the cluster approach is, that this function of four spins around
a plaquette may be written as a linear combination of ten terms consisting of Kronecker
deltas like $\delta_{s_2 , s_3}$ tying together pairs of spins with positive coefficients,
\be
   w = p \left( \begin{array}{l}
     \resizebox{3cm}{!}{\epsfig{file=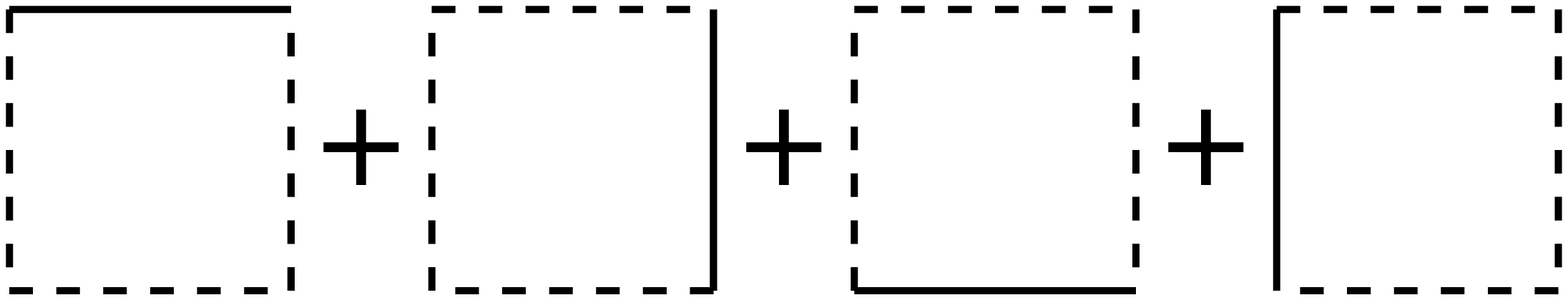}}
   \end{array} \right) + q \left( \begin{array}{l}
     \resizebox{1.4cm}{!}{\epsfig{file=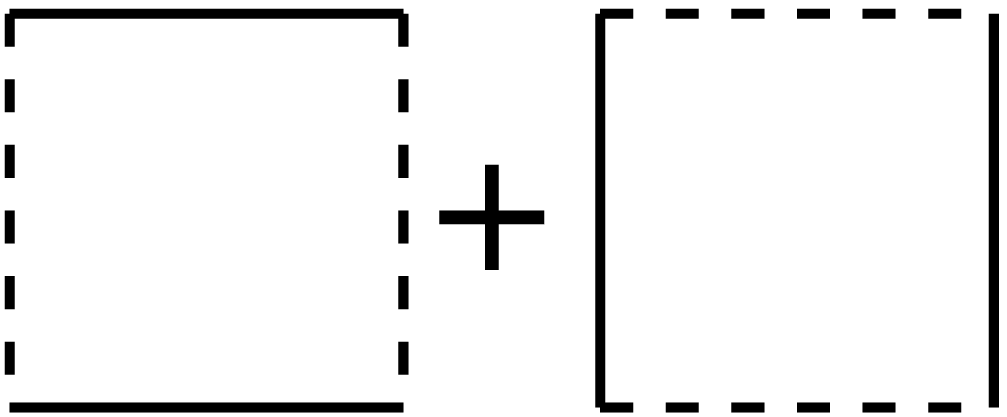}}
   \end{array} \right) + r \left( \begin{array}{l}
     \resizebox{3cm}{!}{\epsfig{file=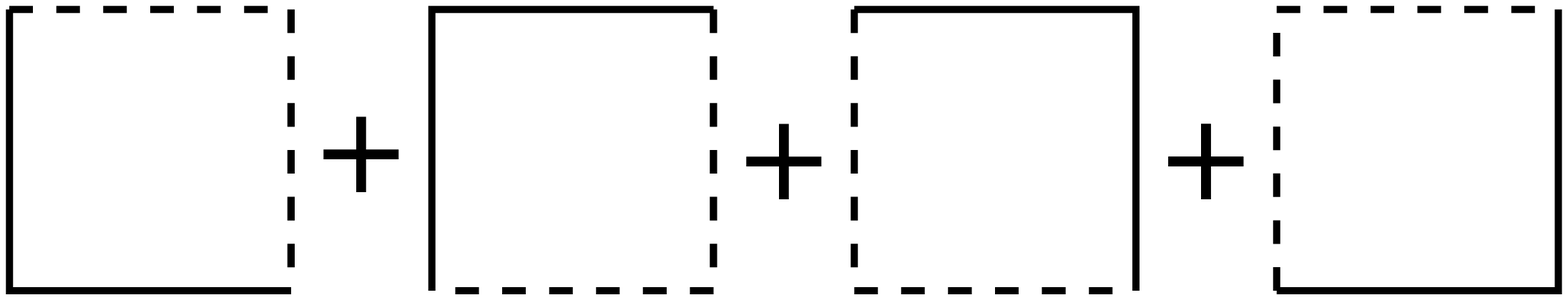}}
   \end{array} \right) .
\label{wdecomp}
\ee
In the above symbolic formula the delta-bonds are represented by the solid lines.
By considering all 16 configurations of the four spins (and the obvious symmetries)
the coefficients
\be
   r = \frac{m (x)}{4}, \hspace{2em} p = \frac{1}{2 \sqrt{2}} - \frac{r}{2},
   \hspace{2em} q = 1 - \frac{1}{\sqrt{2}} + r \hspace{2em} (\tmop{all}
   \geqslant 0 \tmop{ if } m < 2 \sqrt{2}) 
\label{bondprob}
\ee
result. This is a rather direct generalization of the 
Fortuin-Kasteleyn bonds in
\cite{Swendsen:1987ce}, where only just pairs of spins have to be considered.
We now introduce correspondingly 10-valued variables $b$ on the sites of the original
(=plaquettes of the dual) lattice as additional variables.
The partition function is now given
by a sum over both the spin and the new bond field,
\be
   Z_s = \sum_{\{ \text{$b (x) = 1, \ldots, 10$; $s ( \underline{x})$} = \pm   
   1\}}  \prod_{\tmop{plaq}} P_{b (x)} \Delta_{\text{$b (x)$}}
   \left(\begin{array}{cc}
     s_4 & s_3\\
     s_1 & s_2
   \end{array}\right) . 
\ee
Here the factors $\Delta\in{0,1}$ depend on both bonds and spins and contain
the $\delta$ associated with a given $b$-value, while the $P$ are the weights from (\ref{bondprob}).
The decisive trick is now to refrain from exactly summing over ${b}$,
but to alternatingly do Monte Carlo updates to both kinds of variables.
In a first pass, for fixed spins, at each site a new bond is picked among
those allowed by $\Delta$ with relative probabilities given by $P$. This amounts to
a local heatbath, and the $b$-updates at different sites are independent in this step.
With new bonds given we next pick a new spin configuration. All spin configurations on the whole
lattice in this step are either forbidden by $\Delta=0$ or have the {\em same} positive
weight. If one chooses one of the latter at random with equal probability this furnishes a
global heatbath in this step. With the help of percolation cluster search algorithms
which divide the spins into groups (`clusters') that are tied together by the $\delta$
bonds, this becomes possible at a cost of O(volume) only. This is the nonlocal update step
that (almost) eliminates critical slowing down as witnessed by the stars in Fig.\ref{taufig}.

In \cite{Wolff:2007ip} two important generalizations are given that we only mention here.
First, in the step of picking new spins at fixed bonds one may also at the same time consider
to switch among the four possible boundary conditions. One then simulates with `fluctuating'
boundary conditions the ensemble $Z_s^{++}+Z_s^{-+}+Z_s^{+-}+Z_s^{--}$. By reweighting with
a factor depending on the boundary condition as a dynamical variable one can then also
effectively measure at fixed fermion boundary conditions, see (\ref{Zxifix}). This factor
is of fluctuating sign which is a remnant of the fermionic minus sign problem, still
visible at finite volume here. It leads however to no severe cancellations that
cannot be handled. Moreover it even seems possible to construct an improved estimator at
least for some quantities in the reweighted ensemble. The second generalization
allows for $\varphi(x)=2+m(x) <2$, i.e. negative mass. If the simulations studied
for a fermion in the `external' field $m(x)$ here 
are actually embedded to simulate the interacting GN model,
such local masses cannot be avoided. The decomposition (\ref{wdecomp}) with weights
(\ref{bondprob}) requires $m \ge 0$. There is however another decomposition available
in the case of negative mass. It contains both bonds as before and antibonds like
$\delta_{s_1 , -s_2}$ with then altogether 14 terms and positive weights again. Clusters now
also contain tied-together opposite spins, but otherwise the construction of clusters which
are flipped collectively with probability 1/2 is very similar. With local $m(x)$ some
plaquettes may use the one and some the other decomposition.

Both in \cite{Wolff:2007ip} and here only free fermions have been simulated
numerically. In the dimer or Ising form this case in fact does not seem so very special,
but of course from the fermionic realization exact results are available for all quantities.
In the meantime first simulations are running however with interacting $N=2$ and
$N=8$ fermion flavors. First results to be compared with \cite{Korzec:2006hy} look consistent
and the update seems to remain very efficient. Unfortunately the method so far is restricted
to two dimensions. Fermions, and the possibility to `bosonize' them, is known
to be special to this dimensionality. Nevertheless, it seems worth to keep thinking
about `non-determinant' approaches to fermions also for $d=3,4$. 

\noindent Acknowledgement: The author thanks the
Deutsche Forschungsgemeinschaft (DFG)
for support
in the framework of SFB Transregio~9.


\end{document}